\documentclass[12pt]{article}
\usepackage{graphicx}
\begin{document}
\renewcommand{\thefootnote}{\fnsymbol{footnote}}
\begin{titlepage}
\begin{flushright}
hep-th/0402030
\end{flushright}

\vspace{10mm}
\begin{center}
{\Large\bf A Note on  Tachyons in the $D3+{\overline {D3}}$ System}
\vspace{25mm}

{\large
Yan-Gang Miao$^{a,}$\footnote{e-mail address: miao@het.phys.sci.osaka-u.ac.jp},
Harald J.W. M\"uller-Kirsten $^{b,}$\footnote{e-mail address: mueller1@physik.uni-kl.de}
and
Dae Kil Park$^{c,}$\footnote{e-mail address: dkpark@hep.kyungnam.ac.kr}}\\
\vspace{10mm}
${}^a$ {\em Department of Physics, Osaka University,
Toyonaka, Osaka 560-0043, Japan}

\vspace{4mm}
${}^b$ {\em Department of Physics, Technical University of Kaiserslautern, P.O. Box 3049, D-67653 Kaiserslautern,
Germany}

\vspace{4mm}
${}^c$ {\em Department of Physics, Kyungnam University, Masan 631-701, Korea}
\end{center}

\vspace{12mm}
{\centerline {\bf Abstract}}
\vspace{5mm}
\noindent
The periodic bounce of Born-Infeld theory of $D3$-branes is
derived, and the BPS limit of infinite period is discussed
as an example of tachyon condensation. The explicit bounce
solution to the Born--Infeld action is interpreted as an
unstable fundamental string stretched between the
brane and its antibrane.

\vspace{10mm}
\noindent
{\bf Keywords}: nonlinear theories, extended classical solutions, phase transitions

\vspace{3mm}
\noindent
{\bf PACS numbers}: 11.10.Lm, 11.27.+d, 05.70.Fh
\end{titlepage}

\newpage
\setcounter{footnote}{0}
\setcounter{page}{2}

\section{Introduction}

The essential physics and basic equations of  
a system of $D3$ branes
have been worked out in the paper of 
Callan and Maldacena \cite{1}. 
The system of  $D3$ branes is also
one of  the few systems
permitting 
to a considerable extent
explicit calculation  of various aspects.
Since, however, Ref. \cite{1} appeared
before  Sen's work \cite{2}
focussed special interest on tachyons,
the analysis of the $D3$ system in 
that direction is somewhat incomplete.  
In this note we comment on this,  evaluate
an  elliptic integral and consider the
limiting behaviour to the BPS state.

In most models constructed to explore
the behaviour of tachyons  in  $Dp+{\overline {Dp}}$ systems,
two initially unlinked  branes are introduced with their
own respective characteristics such as  gauge fields and tachyons.
The corresponding Lagrangian has  generally not
been written out and hence the 
$Dp+{\overline {Dp}}$ pair then  has not been derived
as a correspondingly explicit solution.   
This also does not permit a straightforward
transition of the initially non-BPS branes
to the BPS state. 
The $D3$-brane theory of Ref. \cite{1}, however,  allows  this,
at least to some extent, to be made  very transparent. 
This is what we demonstrate in the following.
The basic idea behind this is that the  bounce, i.e. periodic 
solution of the equations of motion, has a period which
in the limit of infinite size allows the
solution to become  a topological
object like an instanton or domain wall configuration. 
This process can be described as tachyon condensation.
The periodic solution thus interpolates between
the branes of the brane pair.
Thus one  should find the  periodic
solution.  In fact, a solution of this kind was given in Ref. 
\cite{3} in the case of $D3$ without inclusion of
the electric field, but the solution can be derived more generally.
In the following we derive the full periodic
solution and then demonstrate the transition to  the BPS limit
resulting in  the fundamental string.

\section{Bounces and  the  $D3+{\overline  {D3}}$  system}

We recall briefly from Ref. \cite{1} the points of relevance here.
Taking  the worldbrane gauge field as purely electric and considering
the excitation of only  one transverse coordinate which we here
call $y$ the worldbrane action reduces to
\begin{equation}
S=-\frac{1}{g_3}\int dt\int d^3x\sqrt{(1-{\bf E}^2)(1+\mbox{\boldmath $\nabla$}
y\cdot\mbox{\boldmath $\nabla$}y)
+({\bf E}\cdot{\mbox{\boldmath $\nabla$}y})^2-{\dot y}^2}
\label{1}
\end{equation} 
with $g_3=g(2\pi)^3$, $g$ the string coupling 
in the notation of Ref. \cite{1}.
In Ref. \cite{1} the set of static solutions
 representing strings going between
branes and antibranes is investigated by using spherical symmetry and 
choosing  a single-charge electric solution of the associated
constraints. The equation for  the static
solution $y$ can be integrated to give
\begin{equation}
y(r)=\int^{\infty}_r \, dr\frac{B}
{\sqrt{r^4 - r^4_0}},
\label{2}
\end{equation}
where the length $r_0$ or throat radius (which originates
as a constant of integration) is defined by
\begin{equation}
r^4_0=B^2-A^2.
\label{3}
\end{equation}
The constant $A$ is a parameter related to the electric point 
charge $c_3$, 
from which the electric field originates and is related
to the tension of the fundamental string, i.e.
\begin{equation}
c_3\equiv g\pi=\pi(2\pi)^3g_s \;\; {\rm and } \; \; A=c_3,
\label{4}
\end{equation}
where the constants and parameters are those used in Ref. \cite{1}
which we retain here for easy comparison.
Thus with $A=0$ we eliminate the electric field; the configuration
described by the solution 
then remaining is that usually described as catenoid (see Ref. \cite{3}
and extensive references there).

Thus ignoring the electric field we have  only
the hull, i.e. catenoid part of the $D3$-brane considered
in detail in Ref. \cite{3}.
This is the solution
\begin{equation}
y_c(r)=\pm\int^{\infty}_r dr\frac{{r_0}^{2}}
{\sqrt{r^{4}-{r_0}^{4}}}.
\label{5}
\end{equation}
With  $x=r/r_0$, ${\tilde y}_c(x)
=y_c(r_0x)$, this integral  can be rewritten and integrated as
an elliptic integral so that
\begin{eqnarray}
{\tilde y}_c(x)&=&\pm\int^{\infty}_x\frac{dx}{\sqrt{x^4-1}} 
\nonumber\\
&=&\pm\frac{1}{\sqrt{2}}\bigg[{K}\bigg(\frac{\sqrt{2}}{2}\bigg)  
-{\rm cn}^{-1}\bigg(\frac{1}{x}, \frac{\sqrt{2}}{2}\bigg)\bigg].
\label{6}
\end{eqnarray}
Here ${\rm cn} (u,k)$ is the Jacobian elliptic cosine
function (with the property that ${\rm cn} (u,0)=\cos\,u$)
and $K(k)$ is its quarter period (with the property that
$K(0)=\pi/2$). In these expressions the parameter $k$
is the so-called elliptic modulus which determines 
the elliptic deviation of $2K(k)$ from $\pi$. The function
${\rm cn}^{-1}(u,k)$ is the inverse function in analogy
to its trigonometric counterpart. Properties of
 elliptic integrals and functions can be looked up
in Refs. \cite{4} and \cite{5}. 
We set, $k$ being the elliptic modulus,
$$
a \equiv {K}(k),\qquad \qquad  k=\frac{\sqrt{2}}{2}. 
$$
When $k$ has this value, one has (cf. \cite{4}, formula 111.10)
${K}={K}^{\prime}={K}(k), k^{\prime}=\sqrt{1-k^2}$.
Then, since the variable or argument of the inverse function
${\rm cn}^{-1}$ is $1/x$, we can write the equation
\begin{equation}
x(y)=\bigg[{\rm cn}\bigg(a\mp \sqrt{2}y,k\bigg)
\bigg]^{-1}.
\label{7}
\end{equation}
Suppressing  the elliptic modulus $k$,  we
have
\begin{equation}
x(y)=\frac{1}{{\rm cn }({K}\mp \sqrt{2}y)}.
\label{8}
\end{equation}
Now we use the formula (cf. Ref. \cite{4}, formula 122.03)
${\rm cn}(u+{ K})=-k^{\prime}{\rm sd} u$, where
${\rm sd}u={\rm sn} u/ {\rm dn} u$, the function
${\rm dn}\,u$ being the third of the
three fundamental Jacobian elliptic functions.

Hence
\begin{equation}
\frac{r(y)}{r_0}\equiv
x(y)=\frac{1}{-k^{\prime}{\rm sd}(\mp \sqrt{2}y)}=
\frac{{\rm dn}(\mp \sqrt{2}y)}{-k^{\prime}{\rm sn}(\mp \sqrt{2}y)}
=\frac{{\rm dn}(\sqrt{2}y)}{\pm k^{\prime}{\rm sn}(\sqrt{2}y)}.
\label{9}
\end{equation}
The function ${\rm dn}(u)$ is a slowly varying function a little bit
below $1$
(see the figures in  Ref. \cite {5}; for $k^2=1/2$ the function ${\rm dn}(u)$ 
varies between 1 and (approximately) 0.6 for any value of $u$). 
Thus roughly speaking it is almost a nonvanishing
 constant. We can concentrate on ${\rm sn}(u)$.  Depending
on $k$ this varies between a sine function and the hyperbolic tangent.
Thus $x(y)$ has more or less the behaviour of $1/\sin(\sqrt{2}y)$.
We show this schematically in Fig.~1.
\begin{center}
\includegraphics[angle=0,totalheight=8.0cm]{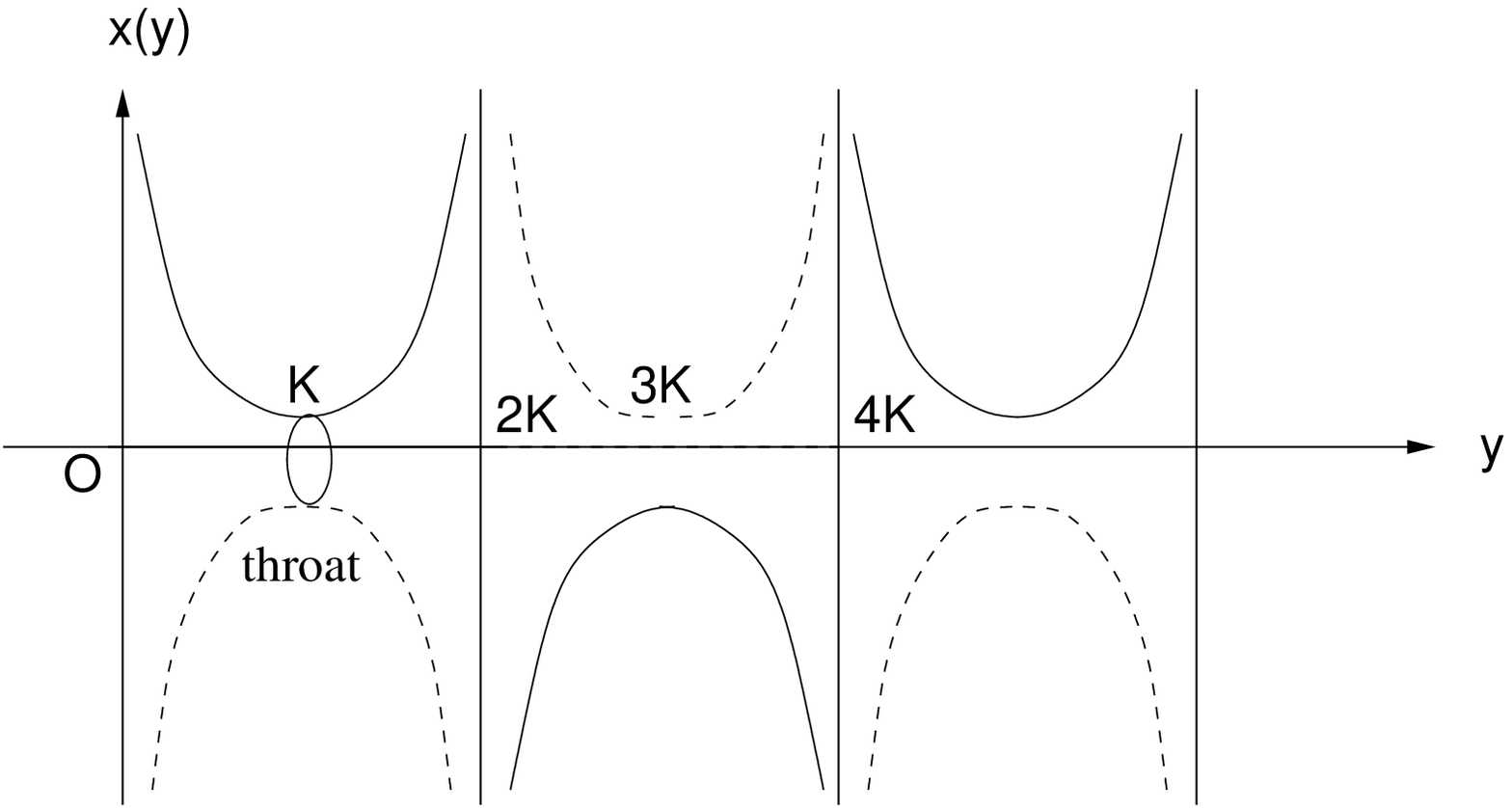}
\end{center}
\centerline{\small Fig.~1  Schematic behaviour  of $x(y)$.}

\noindent
If we rotate the schematic sketch  of $x(y)$ through $90$ degrees, we
obtain a figure like that of Fig.~2(c) of Ref. \cite{1}.
Thus one would like to know whether
one can  make the period so large that
one brane is at infinity and we end up with the
fundamental string.  In the above
the period is  a fixed number which does not
permit this. Comparing Eq.~(\ref{5}) with results
of  Ref. \cite{1}, however, 
we see that the full generalisation of our integral above
for $y_c$, i.e. with inclusion of the electric 
field $E(r)$, is given by Eq.(\ref{2}) above. 
 Thus, in order to incorporate the electric
field we have to make in the above
the replacement
$$
y_c(r)\longrightarrow\frac{X(r){r_0}^{2}}{B}.
$$
The product $g\pi$ is effectively the charge in the $D3$ brane
from which the open string emanates (cf. Eq.~(\ref{4})
above).  For $A=0$ the action
is a maximum, as pointed out in Ref. \cite{1} and as
demonstrated in Ref. \cite{3}.  The configuration
$y_c$ (i.e. that  with $A=0$) is therefore --- in Ref. \cite{1} ---
identified with that at the top of the potential (barrier).
The classical configuration there is usually  described as a sphaleron.
Reference \cite{1} therefore suggests  that there must be
configurations, i.e. solutions below this
maximum, i.e.
when $A\neq 0$, and these are the bounces or --- as some say ---
periodic solitons.  The authors
of Ref. \cite{1} do not evaluate any elliptic integrals.
So if these periodic solutions exist, let's see them and obtain
their period. Replacing $y_c$ by the above expression in terms of
$X$ we have:
\begin{equation}
\frac{r(y)}{r_0}\longrightarrow\frac{r(X)}{r_0}
=\frac{{\rm dn}(\sqrt{2}X{r_0}^2/\sqrt{{r_0}^4+A^2})}
{k^{\prime}{\rm sn}(\sqrt{2}X{r_0}^2/\sqrt{{r_0}^4+A^2})}. 
\label{10}
\end{equation}
We see that for $A=0$ this reduces to the catenoid solution,
i.e. to the configuration described as sphaleron in Ref. \cite{1}.
The solution $r(X)$ yields the periodic bounce.  What is the
period? From the literature (Refs. \cite{4} and \cite{5}) we take that
$$
{\rm sn} (u)\;\; {\rm has \; real \;  period} \;\; 4{K};\;\;
{\rm dn} (u)\;\; {\rm has \; real \; period} \;\; 2{K}.
$$
Here ${\rm sn}(u,k)$ is the Jacobian elliptic
sine function which together with the function ${\rm cn}(u,k)$ satisfies
the relation ${\rm sn}^2u+{\rm cn}^2u=1$, which 
corresponds to 
$\sin^2u+\cos^2u=1$.
Thus $r(X)$ has period  $4{K}$.
 This means, setting
$$
u=\sqrt{2}X{r_0}^2/\sqrt{{r_0}^4+A^2},
$$
that
$$
\frac{{\rm dn}(u+4n{K})}{{\rm sn}(u+4n{K})}=\frac{{\rm dn}(u)}
{{\rm sn}(u)},
$$
and  the period  $P$ in $X$  is
\begin{equation}
P=\frac{4{K}\sqrt{{r_0}^4+A^2}}{\sqrt{2}{r_0}^2}
=\frac{4{K}}{\sqrt{2}}\sqrt{1+\frac{A^2}{{r_0}^4}}.
\label{11}
\end{equation}
Thus, replacing $X$ by $X+P$, one has ${\rm dn}(u)={\rm dn}(u+4{K})$.

Now we wish to calculate the smallest width of the throat, i.e.
the smallest value of  radius $r$ indicated in  Fig.~2.

\vspace{0.6cm}

\begin{center}
\includegraphics[angle=0,totalheight=8.0cm]{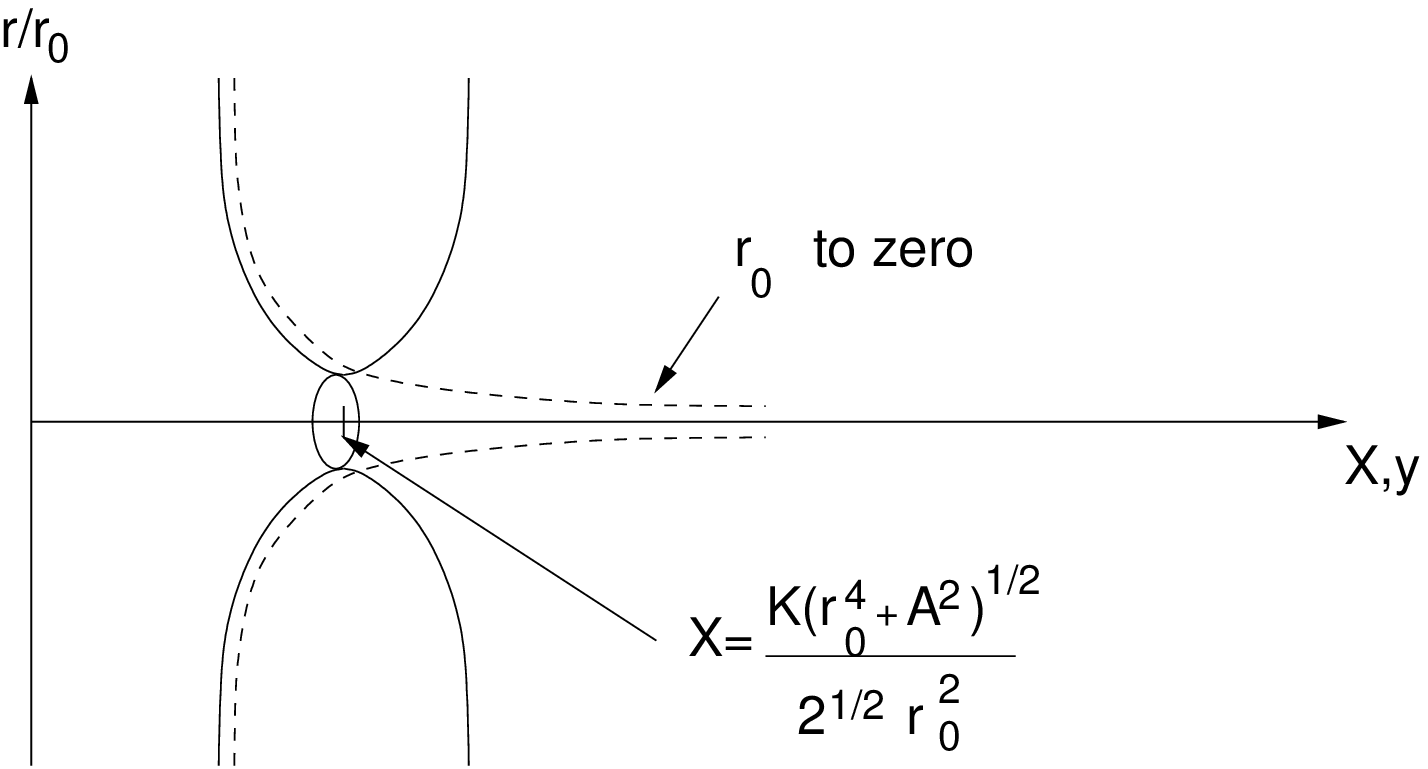}
\end{center}
\centerline{\small Fig.~2 The throat.
 Observe that for $r_0\rightarrow 0$ the minimum}
\centerline{\small wanders off to infinity but only for $A^2\neq 0$.}

\vspace{1.0cm}

\noindent
From the tables of Ref. \cite{5}  we have:
$$
{\rm sn}({K})=1, \;\; {\rm dn}({K})=k^{\prime}.
$$
Inserting this into $r(X)/r_0$, we obtain:
$$
\frac{r({K})}{r_0}=1
$$
with
\begin{equation}
\frac{\sqrt{2}Xr^2_0}{\sqrt{r^4_0+A^2}}={K}, \;\;\;
\frac{{ K}\sqrt{r^4_0+A^2}}{\sqrt{2}r^2_0} =\frac{P}{4}.
\label{12}
\end{equation}
We observe that for $r_0\rightarrow 0$: $X\rightarrow\infty$ and
the antibrane goes to infinity, at the end of the fundamental
string. This is the passage to the supersymmetric
or BPS limit, as discussed in Refs. \cite{1} and 
\cite{3}.  Thus we interpret the spike
carrying the electric field (originating from the
charge at $r=0$)  as the open  string
emanating from one brane and heading towards the other.
This suggests to identify the
fluctuations about the brane-antibrane pair,
i.e. about the periodic solution, with the fluctuations
of the open string, after all, it is the
low energy fluctuations which replace the open string in
our considerations here.  Hence it is not only
the negative-eigenvalue or ground-state mode
which has to be identified with the
tachyon, but correspondingly also the
zero-eigenvalue or first-excited-state mode which has to be identified with
(here) the electric field.

The next step would be to look at the equation of small
fluctuations about the periodic solution above.  Then, by
equating to zero the electric field or charge, one should
arrive at the fluctuation equation about the catenoid ---
this is the equation with differential operator given by
Eq.~(33) in Ref.\cite{3} --- and by imposing the BPS conditions
(Eq.~(64) of Ref.\cite{3}) one should obtain the fluctuation
equation about the BPS solution, i.e. Eq.~(74) of Ref.\cite{3}.
The procedure to derive this equation would be that of Ref.\cite{3}
by following the steps after Eq.~(62), however without imposing
immediately the BPS conditions there given by Eq.~(64).
This seems to require some longer but straightforward calculations.
We skip these here and instead ---
after the following discussion ---  present plausibility
arguments based on analogous but easily solvable 
examples.

Thus we arrive at the following picture.  The brane-antibrane pair
with the brane and antibrane
a finite distance apart is a periodic solution of
the Euler-Lagrange equations obtained from the Born-Infeld
Lagrangian. Since  
 in the  two limiting cases  of BIon (no field $y$)
 and catenoid (no gauge field) the solution is either
stable or unstable as discussed in detail in Ref. \cite{3}, 
in the general case
 the brane-antibrane pair is unstable and hence
the equation of small fluctuations about this
solution must possess a negative eigenvalue, as is
wellknown (recall that a periodic solution has the
shape of a squeezed instanton--antiinstanton pair which
pushes one of these to infinity when the period is
made infinite, thereby removing the negative mode).
Arguments of translation invariance imply that there must
also be a zero mode, i.e. a solution of the small fluctuation
equation with eigenvalue zero.  These solutions describe
the amplitudes of the fluctuations in the direction   
perpendicular to the $D3$  branes. 
We recall that at low energies the effect of an open string,
i.e. its fluctuations, is approximated by its lowest eigenmodes,
and these are the tachyon and gauge field modes with
negative mass-squared and zero mass respectively.
Thus it is suggestive to
 identify these lowest  oscillations of the open string
with the lowest fluctuations about the brane-antibrane
pair  in the above sense. A scalar  tachyon field $\phi$
was not inserted into the theory originally, however,
the tachyon mode arising here can be considered as a 
remnant of  such a  field as a function of  one
co-dimension orthogonal to the brane.  A similar though different
consideration may apply to the gauge field
and the zero mode. These aspects have been discussed
in Ref. \cite{6}.

Since we do not enter here into a detailed study
of the small fluctuation equation and its spectrum,
we restrict ourselves --- as stated above --- to
 arguments of analogy. The simplest example
of a nonperiodic  bounce is obtained for the (admittedly unphysical)
cubic potential as demonstrated in Ref. \cite{7}
and the fluctuation equation is the equation
with the well-known P\"oschl--Teller potential
(of $1/{\cosh}^{2}x$ type with $x \in \mathbf{R}$).
 The solution (i.e. ``zero mode'') associated
with the eigenvalue zero is, as usual
for reasons of translational invariance, given
by the derivative of the classical configuration
and when this is an odd
function   the 
ground state has a  negative eigenvalue.  
These   general properties of bounces   
have been discussed by Coleman \cite{8}.
In fact Coleman has given general arguments
to the effect that a bounce is associated
with one  physical negative  eigenvalue of
the fluctuation equation.

Periodic configurations arise in periodic
cases.
Thus in Ref. \cite{9} the very  popular
examples of double-well potential, inverted double-well
potential and cosine potential were considered on
a circle and the nontopological,
periodic instanton-like  configurations
were derived. Very naturally these solutions  were
found to be Jacobian elliptic functions like
the function ${\rm sn} (u,k)$
in the case of the double well potential.
Such a solution reduces in the limit
of $k=1$ to the well-known topological  kink solution,
i.e. ${\rm tanh} x$, and therefore is not
a topological vacuum for other values of $k$.
The effect on the quarter period $K(k)$
of allowing $k$ to vary from $0$ to $1$ 
is to vary from $\pi/2$ to $K(1)=\infty$. Thus
the periodic solution varies from a trigonometric
form to  the hyperbolic tangent.
Considering now 
fluctuations about these configurations, one
obtains --- as explicitly demonstrated in Ref. \cite{8}
--- as fluctuation equation the wellknown Lam${\acute e}$
equation whose solutions can be looked up in books 
like Ref. \cite{10}.
In each case these equations have several solutions
with negative, zero and positive eigenvalues, some of
which merge together in the limit $k\rightarrow 1$
with the disappearance of negative modes in 
topological cases.

\section{Concluding remarks}
As noted, in the   $D3+{\overline  {D3}}$ system discussed above
the tachyon field and tachyon potential do not appear
explicitly, instead only indirectly through  the negative mode.
Models dealing with explicit tachyon potentials,
such as the open string field theory models
of Ref. \cite{6},
permit the explicit derivation of the fluctuation modes.
In these cases the tachyon potential is either of the
unbounded cubic type (in the bosonic theory)   
or of double-well type (in the superstring theory).
The fluctuation spectra of these potentials
have been considered in detail in Refs. \cite{7} and
\cite{9} respectively.

We know from other  studies in Ref.~\cite{9} that the negative
eigenvalue mode merges with the zero mode in the limit
in which the period of the periodic solution or bounce
becomes infinite, i.e. when the periodic solution becomes
a soliton. In effective tachyon scalar superstring theories
as discussed for instance in Ref. \cite{6} the calculations 
are essentially those familiar from soliton theory, the
tachyon scalar field depending on only one co-dimension
which is orthogonal to the remaining spatial coordinates of
the original $Dp$-brane.  With the tunneling
through the central hump of the double-well potential, the
original perturbation vacuum energy is
lowered to the real ground state, which is characterised
by one of the two topological and hence
BPS states provided by the kink or domain wall solution.
The final state thus reached by ``tachyon condensation'' then
is the $D(p-1)$-brane which carries a topological charge.
In this process the period of the periodic solutions becomes infinite.  
This seems to be the process also described in Ref. \cite{11}, where
periodicity arises by putting the domain wall on
a circle as demonstrated explicitly in several
cases in Ref. \cite{9}.  
A situation analogous to the $D3$ case considered here can
be seen in the case of $D2$-branes, although there the
reverse process has been discussed,
i.e. that of strings tunneling to branes, as in Refs. \cite{12}
and \cite{13}.

\vspace{1cm}
\noindent
{\bf Acknowledgements} 

D.K.P. acknowledges support by the Korea
Research Foundation under Grant No. KRF-2003-015-C00109.

\end{document}